\documentclass[aps,prd,groupedaddress,nofootinbib]{revtex4-1}

\usepackage{dcolumn}
\usepackage{bm}

\usepackage{graphicx}
\usepackage{amsmath}
\usepackage{here}

\usepackage{here}
\usepackage[dvips]{epsfig}
\def\beq{\begin{equation}}
\def\eeq{\end{equation}}
\def\bea{\begin{eqnarray}}
\def\eea{\end{eqnarray}}
\def\bq{\begin{quote}}
\def\eq{\end{quote}}

\def\nnb{\nonumber}
\def\ga{\left(}
\def\dr{\right)}

\def\lrar{\Longrightarrow}
																
\def\nnb{\nonumber}

\def\la{\langle}
\def\ra{\rangle}
\def\nin{\noindent}
\def\ba{\vspace*{-0.2cm}\begin{array}}
\def\ea{\end{array}\vspace*{-0.2cm}}

\def\b{$\bullet~$}
\def\d{$\diamond~$}

\def\als{\alpha_s}

\def\gg2{\la\alpha_s G^2 \ra}
\def\gg3{g^3f_{abc}\la G^aG^bG^c \ra}
\def\ggg4{\la\als^2G^4\ra}

\def\gg{\lag g^{2}_{s} G^2 \rag}
\def\ggg{\lag g^{3}_{s}G^3\rag}

\usepackage{amsmath}
\usepackage{slashed}
\usepackage{color}

\begin{document}

\title{Laplace Sum Rules in Quantum ChromoDynamics}
\thanks{Invited contribution to the book : The Laplace Transform and its Applications edited by Nova Science Publishers, New-York.}
\author{Stephan Narison
}
\affiliation{Laboratoire
Univers et Particules de Montpellier (LUPM) \\ CNRS-IN2P3, and University of 
 Montpellier, France \\
and \\
Institute of High-Energy Physics of Madagascar (iHEPMAD)\\
University of Ankatso, Antananarivo, Madagascar}
\email[Email address:~] {snarison@yahoo.fr} 


\begin{abstract}
\noindent
We  shortly review some applications of the (inverse) Laplace (LSR) transform sum rules in Quantum ChromoDynamics (QCD) for extracting the fundamental QCD parameters (coupling constant $\alpha_s$, quark and gluon condensates) and the hadron properties (masses and decay constants). Links of LSR to some other forms of QCD spectral sum rules are also discussed. As prototype examples, we discuss in detail the $\rho$ and $\pi$ meson sum rules. \\
\\
{\bf Keywords :} {\it QCD spectral sum rules, QCD coupling $\alpha_s$ quark masses and condensates,  
\hspace*{2.3cm}Hadron masses and couplings.}
\end{abstract}
\maketitle

\section{Introduction}
\vspace*{-0.2cm}
Different spectral sum  rules based on dispersion relations called current algebra sum rules have been used before QCD\,\cite{FURLAN}. Among these, there are the Weinberg\,\cite{WEIN} and Das-Mathur-Okubo (DMO)\,\cite{DMO} sum rules which assume the asymptotic realizations of chiral $SU(2)_L\otimes SU(2)_R$ and $SU(2)_V$ flavour symmetries to derive constraints on the vector and axial-vector mesons masses and couplings.  One of the famous predictions from the Weinberg sum rules is the mass relation:
\beq
M_{A_1}\simeq \sqrt{2}M_\rho,
\eeq
which is reproduced within the errors by the data.  

Within the advent of QCD, corrections to these sum rules have been studied\,\cite{FNR,DMOQCD,GMONARISON}  in 1979-80. At about the same time, 
Shifman-Vainshtein and Zakharov (hereafter referred as SVZ)\,\cite{SVZa,ZAKA}\,\footnote{For reviews, see e.g. the books\,\cite{SNB1,SNB2}, the reviews in\,\cite{SNB3,RRY,BERTa,REV} and the recent ones in Refs.\cite{SNREV22,SNREV21,SNbc15}.}, have introduced the QCD spectral sum rules 
which are the improvement of the usual dispersion relation obeyed by e.g. the hadron two-point function:
 \beq
\hspace*{-0.6cm} \Pi_H(q^2)=i\hspace*{-0.1cm}\int \hspace*{-0.15cm}d^4x ~e^{iqx}\la 0\vert {\cal T} {\cal O}_H(x)\ga {\cal O}_H(0)\dr^\dagger \vert 0\ra=\int_{t_>}^\infty \frac{dt}{t-q^2-i\epsilon}\frac{1}{\pi}{\rm Im}\Pi(t)+\cdots,
 \label{eq:2-point}
 \eeq
where $\cdots$ mean subtraction constants polynomial in the momentum $q^2\equiv -Q^2<0$. 

${\cal O}_H(x)$ is a generic notation for a hadronic current:

-- Quark bilinear local current $\bar \psi_1\Gamma_{12}\psi_2$ for mesons. 
$\Gamma_{12}$ is any Dirac matrices which specify the quantum numbers of the corresponding meson state (and its radial excitations),

-- Quark trilinear local current $\psi_1\psi_2\psi_3$ for baryons,

--   Four-quark $(\bar \psi_1\Gamma_{12}\psi_2)(\bar \psi_3\Gamma_{34}\psi_4)$ or diquark anti-diquark $(\bar \psi_1\Gamma_{12}\bar\psi_2)(\psi_3\Gamma_{34}\psi_4)$ local current for molecules or tetraquark states, 

--  Pentaquark $\psi_1\psi_2\psi_3\psi_4\psi_5$ states,

-- Gluon local currents $G^2, G^3\cdots$ for gluonia/glueball states,

-- Quark-gluon local currents $\bar\psi_1G\psi_2, \bar\psi_1 G^2\psi_2\cdots $ for hybrid states.

In the case of a $\bar \psi_1\psi_2$ meson bound state built from a quark $u$ and the anti-quark $\bar d$, the current reads:
\beq
{\cal O}_H(x)=\bar u \Gamma d (x),
\eeq
where $\Gamma$ denotes any Dirac matrices which specify the quantum number of the associated hadron. $\Gamma = \gamma_u,\gamma_\mu\gamma_5,\gamma_5,1$ for the vector, axial-vector, pseudoscalar  and scalar currents which 
correspond respectively to the $\rho,A_1,\pi$ and scalar quarkonium mesons. 
\vspace*{-0.3cm}
\section{The SVZ Borel/Laplace sum rules (LSR)}
\begin{figure}[hbt]
\begin{center}
\includegraphics[width=12cm]{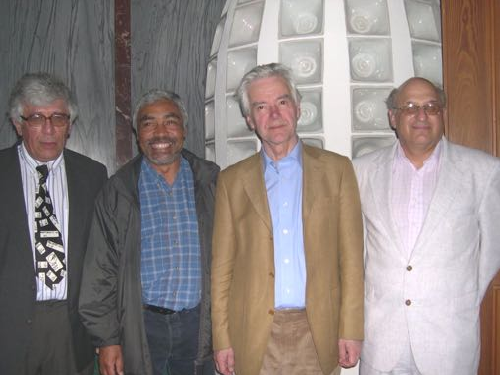}
\vspace*{-0.15cm}
\caption{\footnotesize  Left to right : Arkady Vainshtein, SN, Valya Zakharov, Mikhael Shifman at the Munich conference (2006). }
\label{fig:svz}
\end{center}
\vspace*{-0.5cm}
\end{figure}

\vspace*{-0.20cm}
 Besides the Shifman-Vainshtein-Zakharov (SVZ) (see Fig.\,\ref{fig:svz}) theoretical improvement of the perturbative QCD expression within the Operator Product Expansion (OPE) where quark and gluon  condensates of higher and higher dimensions are supposed to approximate the non-perturbative QCD contributions which we shall discuss in the next section,  the phenomenological success of QCD (spectral) sum rules or SVZ sum rules  also comes from the improvement of the usual dispersion relation which is the bridge between the high-energy QCD expression and the measurable spectral function at low energy.   In the example of the light  isovector ($I=1$) vector current:
 \beq
 J^\mu_H(x)=\frac{1}{2}{[}: \bar\psi_u\gamma^\mu\psi_u-\bar\psi_d\gamma^\mu\psi_d:{]}.
 \eeq
  the spectral function is related to the $e^+e^-\to$ Hadrons total cross-section via the optical theorem:
 \beq
 R_{ee}\equiv\frac{\sigma(e^+e^-\to\,{\rm Hadrons)}}{\sigma(e^+e^-\to\mu^+\mu^-)}=\ga\frac{3}{2}\dr 8\pi\, {\rm Im} \Pi_H(t).
 \eeq
\vspace*{-0.25cm}
\subsection{The form of the Laplace sum rules}
This improvement has been achieved by working with large number $n$ of derivatives and large value of the momentum transfer $Q^2$ but taking their ratio $\tau\equiv n/Q^2$ finite leading to the so-called SVZ Borel/Laplace or Exponential sum rules (LSR) and their ratios,\footnote{Non-relativistic version of this sum  rule has been discussed by\,\cite{BELLa,BERTa}, while the inclusion of the PT $\alpha_s$ correction to the QCD expression has shown that it has the property of an inverse Laplace transform in SNR\,\cite{SNR} though the name LSR.} :
\bea
{\cal L}_0^c(\tau,\mu)&\equiv&\lim_ {\begin{tabular}{c}
$Q^2,n\to\infty$ \\ $n/Q^2\equiv\tau$
\end{tabular}}
\frac{(-Q^2)^n}{(n-1)!}\frac{\partial^n \Pi}{ ( \partial Q^2)^n}
=\int_{t>}^{t_c}dt~e^{-t\tau}\frac{1}{\pi} \mbox{Im}\Pi_H(t,\mu)~,
\nnb\\
 {\cal R}^c_{10}(\tau)&\equiv&\frac{{\cal L}^c_{1}} {{\cal L}^c_0}= \frac{\int_{t>}^{t_c}dt~e^{-t\tau}t\, \mbox{Im}\Pi_H(t,\mu)}{\int_{t>}^{t_c}dt~e^{-t\tau} \mbox{Im}\Pi_H(t,\mu)},~~~~
\label{eq:lsr}
\eea
where $\tau$ is the LSR variable, $t>$   is the hadronic threshold.  Here $t_c$ is  the threshold of the ``QCD continuum" which parametrizes, from the discontinuity of the Feynman diagrams, the spectral function  ${\rm Im}\,\Pi_H(t,m_Q^2,\mu^2)$.  

\d Thanks to the exponential weight and for moderate values of $\tau$, the previous sum rule improvements enhance the low energy contribution to the spectral integral which is accessible experimentally. 

\subsection{The minimal duality ansatz (MDA)} 
In the often case where the data
on the spectral function are not available, one usually parametrizes it via the minimal duality ansatz :
\vspace*{-0.2cm}
\bea
\hspace*{-0.25cm}
\frac{1}{\pi}\mbox{ Im}\Pi_H(t)\simeq  f^2_HM_H^{2d}\delta(t-M^2_H)
  +
  ``\mbox{QCD continuum}" \theta (t-t_c),
\label{eq:duality}
\eea
 in order to predict the masses and couplings of the lowest ground state and in some case the ones of its radial excitations.  $d$ depends on the dimension of the current, $f_H$ is the hadron decay constant  normalized as $f_\pi=$ 132 MeV. Its accuracy has been tested in various light and heavy quark channels $e^+e^-\to \rho,J/\psi, \Upsilon,\dots$ where complete data are available\,\cite{SNB1,SNB2} and in the $\pi$-pseudoscalar channel where an improved parametrization of the $3\pi$ channel within chiral perturbation theory has been used\,\cite{BIJNENS}. Within a such parametrization, the ratio of sum rules is used to extract the mass of the lowest ground state as it is equal to its square. 
\subsection{Hadron masses from the ratio of moments} 
 Within the MDA parametrization, the ratio of sum rules ${\cal R}_n^c(\tau)$  is used to extract the mass of the lowest ground state as it is equal to its square:
 \beq
 M_H^2\simeq {\cal R}^c_{10} (\tau). 
 \eeq
  However, this analysis cannot be done blindly without studying / checking the  moments ${\cal L}_{0,1}$ which in some cases can violate positivity of the specral integral for some values of the sum rule variables ($\tau$)  though their ratio can lead to a positive number identified with the hadron mass squared.  
\subsection{Hadron mass-splittings from the double ratio of sum rule (DRSR) \,\cite{DRSR}\,}
The hadron $H$ and $H'$ mass-splitting, like the one due to $SU(3)$ breakings, can be derived from the quantity:
\beq
r_{H'/H}\equiv \frac{{\cal R}_{10}^c(\tau')\vert_{H'}}{{\cal R}_{10}^c(\tau)\vert_{H}}\simeq \frac{M_{H'}^2}{M_{H}^2},
\eeq
provided that the optimal value: $\tau'_0\simeq \tau_0$. A similar quantity can be used for the heavy quark moments.

 \vspace*{-0.3cm}
\section{Link of LSR to some other sum rules}
\vspace*{-0.15cm}
\subsection{Local duality Finite Energy Sum Rules (FESR)}
It has the form:
\beq
 {\cal F}^c_n(t_c,\mu)=\int_{t>}^{t_c}\hspace*{-0.cm}dt\,t^n\frac{1}{\pi} \mbox{Im}~\Pi_H(t,\mu)~,
\eeq
where $n\geq 0$ is the degree of the sum rule. This sum rule has been used before QCD in Refs.\,\cite{LOGUNOV,SAKURAI} and within QCD in Refs.\,\cite{KRASNIKOV, KATAEV,LARIN}.

This sum rule can be derived from LSR by using a small $\tau$-expansion and by matching for a given $\tau$ its QCD and experimental side. 

FESR can be used to fix the value of the QCD continuum threshold $t_c$ from its dual constraint with the mass and decay constant of the ground state\,\cite{FESR1,FESR2}. 
Contrary to LSR, FESR emphasizes the role of higher masses radial excitations to the integral. Therefore, it requires a good control of the high-energy part of the spectral function.
\subsection{Gaussian Sum Rules }
It  has the form\,\cite{FESR1,FESR2}:
\beq
{G}^c_n(s,\sigma,\mu)=\frac{1}{\sqrt{4\pi\sigma}}\int_{t>}^{t_c}dt~e^{-\frac{(t+s)^2}{4\pi}}\frac{1}{\pi} \mbox{Im}\Pi_H(t,\mu)~,
\eeq
for a Gaussian centered at $s$ with a finite width resolution $\sqrt{4\pi\sigma}$. It has been also shown in Ref.\,\cite{FESR1,FESR2} that the LSR can be derived from the Gaussian sum rule using the $\zeta$-prescription.  

\section{The SVZ - Operator Product Expansion (OPE)\label{sec:ope}}
\d According to SVZ, the Right Hand Side (RHS) of the two-point function can be evaluated in QCD within the Operator Product Expansion (OPE) provided that $ \Lambda^2\ll Q^2\equiv -q^2, m_Q^2$ where $\Lambda\simeq (342\pm 8)$ MeV is the QCD scale for 3 flavours and $m_Q$ is the heavy quark mass. In this way, it reads\,:
\beq
4\pi^2\Pi_H(q^2,m_Q^2,\mu)=\sum_{D=0,1,..}\hspace*{-0.25cm}\frac{C_{2D}(q^2,m_Q^2,\mu)}{Q^{2D}}\la O_{2D}(\mu)\ra~, 
\label{eq:ope}
\eeq
where, in addition to the usual perturbative QCD contribution (unit operator), one has added the ones due to non-perturbative gauge invariant quark and gluon condensates $\la O_{2D}(\mu)\ra$ having a dimension $2D$ which have been assumed to  parametrize approximately the not yet under good control QCD confinement.  $C_{2D}$ are separated calculable Wilson coefficients in Perturbative (PT) QCD:

\subsection{The usual perturbative (PT) contribution}
It corresponds to $D=0$ while the quadratic quark mass corrections enter via $D=1$. 
\subsection{The quark and gluon condensates}
They  contribute through the  OPE. Up to $2D=6$, they are successively the :

-- $2D=4$ quark and gluon $ m\la\bar\psi\psi\ra$ and $ \la \alpha_s G^2\ra$, 

-- $2D=5$ mixed quark-gluon: $\la\bar\psi\sigma^{\mu\nu}\frac{\lambda_a}{2}G_{\mu\nu}^a\psi\ra$

-- $2D=6$ four-quark and three-gluon:  $\la\bar\psi  \Gamma_1\psi \bar\psi \Gamma_2\psi\ra$ and  $ \la g^3f_{abc} G^a_{\mu\rho}G^{b,\rho}_{\nu}G^{c,\nu}_\rho\ra$. 

-- and so on. 
\subsection{The $2D=4$ condensates}
The quark condensate $m\la\bar\psi\psi\ra$ and the part of the trace of the energy-momentum transfer\,: $\theta^\mu_\mu\vert_g\equiv m\gamma \la\bar\psi\psi\ra +(1/4)\beta \la  G^a_{\mu\nu}G_a^{\mu\nu}\ra$ are known to be subtraction $\mu$-independent where $\gamma,\, \beta$ are the quark mass anomalous dimension and Callan-Symanzik $\beta$-function.  

-- The $\la\bar\psi\psi\ra$ condensate can be deduced from the well-known Gell-Mann, Oakes, Renner relation\,\cite{GMOR}:
\vspace*{-0.15cm}
\beq
(m_u+m_d)\la\bar\psi_u\psi_u+\bar\psi_d\psi_d\ra  = -m_\pi^2f_\pi^2,
\eeq
\vspace*{-0.15cm}
once the running light quark mass is known ($m_\pi, f_\pi =132$ MeV are the pion mass and decay constant) or directly from light baryon sum rules\,\cite{DOSCHSN}. With the value of the running $u,d$ quark masses given in Table\,2 in Ref.\,\cite{SNREV22}, one can deduce the value of  $\la\bar\psi\psi\ra$ in this Table\,2. 

-- The original value of the $2D=4$ gluon condensate $\la \alpha_s G^2\ra=0.04$ GeV$^4$ of SVZ\,\cite{SVZa,RRY} has been claimed\,\cite{BELLa,BERTa} from charmonium sum rules and  Finite Energy Sum Rule (FESR) for $e^+e^-\to I=1$\, Hadrons\,\cite{FESR1,FESR2} to be underestimated. Recent analysis from $e^+e^-\to$   Hadrons, $\tau$-decay and charmonium confirm these claims with the  present updated  average
(see the different determinations in Table 1 of\,\cite{SNparam})\,:
\vspace*{-0.15cm}
\beq
\la \alpha_s G^2\ra=(6.49\pm 0.35)\times 10^{-2}~{\rm GeV}^4. 
\label{eq:g2}
\eeq
\vspace*{-0.15cm}

\subsection{The $2D=5$ quark-gluon mixed condensate} 
It is usually parametrized as $g\la\bar \psi G\psi\ra=M_0^2\la \bar \psi\psi\ra$. It  mixes under renormalization and runs as $(\alpha_s)^{1/(6\beta_1)}$ in the chiral limit $m=0$\,\cite{SNTARRACH}. The scale :
\vspace*{-0.15cm}
\beq
M_0^2=0.8(2)~{\rm GeV}^2,
\eeq
\vspace*{-0.15cm}
has been phenomenologically estimated from light baryons\,\cite{IOFFE,DOSCH,PIVOm} and heavy-light mesons\,\cite{SNhl} sum rules.

\subsection{The $2D=6$ four-quark condensates} 
The renormalization of  dimension-six  condensates has been studied in\,\cite{SNTARRACH} where it has been shown that these condensates mix under renormalization and then cannot be compatible with the vacuum saturation assumption used by SVZ.  Its phenomenological estimate from $\tau$-like decays\,\cite{SNTAU}, $e^+e^-\to$ Hadrons data\,\cite{LNT}, $\tau$-decay\,\cite{SOLA}, FESR \,\cite{FESR1,FESR2}  and baryon\,\cite{DOSCH} sum rules, leads to the average\,:
\vspace*{-0.15cm}
\beq
\rho \alpha_s\la\bar\psi\psi\ra^2\simeq 5.8(9)10^{-4}\,{\rm GeV}^6,
\label{eq:d6}
\eeq
\vspace*{-0.15cm}
indicating a huge violation of the vacuum saturation or factorization:
\beq
\alpha_s\la\bar\psi\psi\ra^2\vert_{fac} = 1.0(9)\times 10^{-4}\,{\rm GeV}^4,
\label{eq:fac}
\eeq
by a factor $\rho$ about 5.8.
 
 \d Fixing the ratio $\la \alpha_s G^2\ra/\rho \alpha_s\la\bar\psi\psi\ra^2= 106(12)$ GeV$^{-2}$ as quoted in Ref.\,\cite{SND21} which reduces the analysis to a one-parameter fit, one deduces from LSR\,\cite{SNTAU}:
 \vspace*{-0.45cm}
 \beq
 \la \alpha_s G^2\ra=  (6.1\pm 0.7)10^{-2}~{\rm GeV}^6,
 \eeq
 \vspace*{-0.15cm}
which shows the self-consistency of the previous numbers.  Some other consistency tests can be found in\,\cite{SNTAU}.

\subsection{The $2D=6$ $g^3f_{abc}\la G^aG^bG^c \ra$ condensate}
 It does not mix under renormalization and behaves as $(\alpha_s)^{23/(6\,\beta_1)}$\,\cite{SNTARRACH}, where $\beta_1=-(1/2)(11-2n_f/3)$ is the first coefficient of the $\beta$-function and $n_f$ is number of quark flavours. 
The first improvement of the estimate of the $g^3f_{abc}\la G^aG^bG^c \ra$ condensate was the recent direct determination of 
the ratio of the dimension-six gluon condensate $\la g^3f_{abc} G^3\ra$ over the dimension-four one $\la\alpha_s G^2\ra$ using heavy quark sum rules with the value\,\cite{SNcb1}:
\vspace*{-0.15cm}
\beq
\rho_G\equiv \la g^3f_{abc} G^3\ra/ \la \alpha_s G^2\ra=(8.2\pm 1.0)~{\rm GeV}^2,
\label{eq:rcond}
\eeq
\vspace*{-0.15cm}
which differs significantly from the instanton liquid model estimate\,\cite{NIKOL2,SHURYAK,IOFFE2}. This result may question the validity of the instanton result. 
Earlier lattice results in pureYang-Mills found:  $\rho_G\approx 1.2$ GeV$^2$\,\cite{GIACO} such that it is important to have new lattice results for  this quantity. Note however, that the value given in Eq.\,\ref{eq:rcond} might also be an effective value of all unknown high-dimension condensates not taken into account in the analysis of \,\cite{SNcb1} when requiring the fit of the data by the truncated OPE.  We have seen  in some examples\,\cite{SNREV21,SNREV22} that the effect of  $ \la g^3f_{abc} G^3\ra$ is a small correction at the stability  region where the optimal results are extracted.  
 
\subsection{Higher dimension condensates}
 Usually, the truncation of the OPE up to $2D=6$ provides enough information for extracting the masses and couplings of the ground state hadrons with a good accuracy. In many papers, some classes of higher dimension terms up to 2D=12 !  are added in the OPE. However, it is not clear if such  term gives the dominant contributions compared to the omitted ones having the same dimension.  The size of these high-dimension condensates is not also under a good control due to the eventual violation of the vacuum saturation used for their estimate. 
\subsection{Beyond the SVZ-OPE}
Different contributions beyond the standard SVZ-OPE (tachyonic gluon mass, small size instantons, duality violation) can be consulted in the recent reviews\,\cite{SNREV21,SNREV22}.
\section{The different terminologies of the SVZ sum rules}
\subsection{The origin of the name: Borel sum rule from the OPE}
Applying the operator in Eq.\,\ref{eq:lsr} to the QCD expression in Eq.\,\ref{eq:ope}, the OPE of the 1st moment becomes:
\beq
{\cal L}_0(\tau,m_q^2,\mu)= \sum_{D=0,1,..}\hspace*{-0.25cm}C_{2D}(\tau,m_q^2,\mu)\frac{\tau^{D}}{(D)!}\la O_{2D}(\mu)\ra~,
\eeq
where the appearance of the $(D)!$ factor in the OPE indicates the property of the Borel transform for the moment sum rule thus the name {\it Borel sum rule}.   For this purpose, we recall that if we consider a function $f(x)$, its Borel transform  is:
\beq
\tilde f(\lambda)=\frac{1}{2\pi\,i }\int_{c-i\infty}^{c+i\infty}e^{\lambda/x}f(x)\,x\,d(1/x),
\eeq
where the integration contour runs to the right of all the singularities of the function $f(x)$. Then, the inverse Borel transform reads:
\beq
xf(x) = \int_0^\infty f(x)\,e^{-\lambda/x} d\lambda.
\eeq
Therefore, if 
\beq
f(x)=a_0+a_1x+a_2x^2+\cdots+a_kx^k+\cdots,
\eeq
 then :
 \beq
 \tilde f(\lambda)=\frac{a_0}{0!}+\frac{a_1}{1!}\lambda+\frac{a_2}{2!}\lambda^2+\cdots+\frac{a_k}{k!}\lambda^k+\cdots.
 \eeq
 The great advantage of this property is the improved convergence of the OPE for the Borel transform $\tilde f(\lambda)$ compared to the one of the original two-point function $f(x)$. 
\subsection{The origin of the name: Laplace sum rule (LSR) including $\alpha_s^n$ corrections}
\begin{figure}[hbt]
\begin{center}
\includegraphics[width=7.cm]{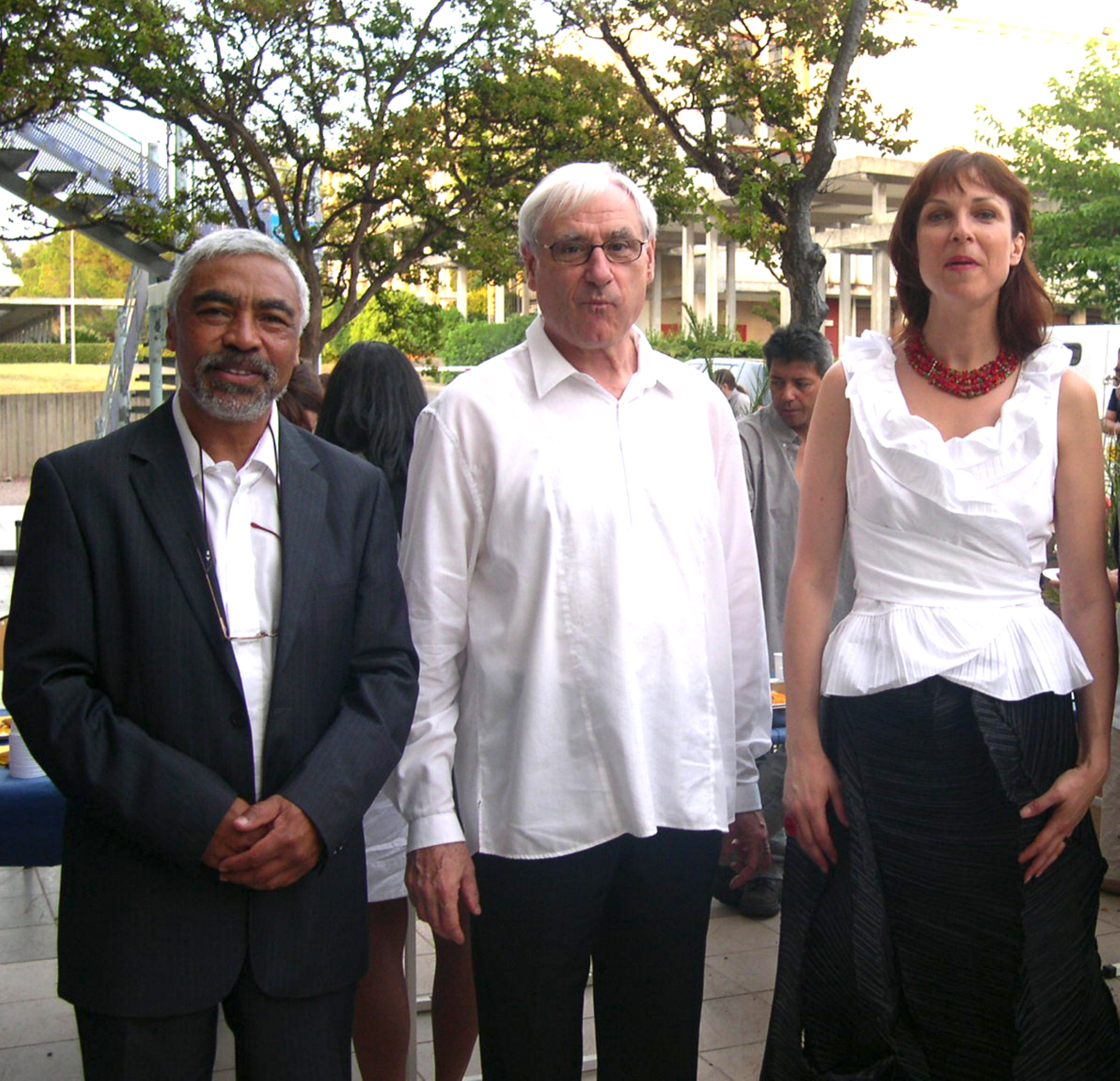}
\includegraphics[width=9.cm]{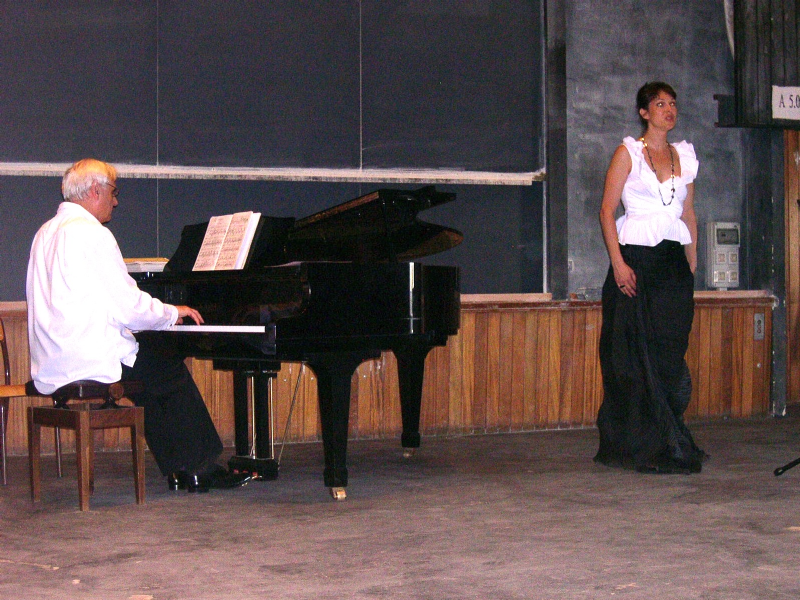}
\vspace*{-0.15cm}
\caption{\footnotesize  Left to right : SN, Eduardo de Rafael, Christine Kattner (soprano singer)  @ QCD 2000 - Montpellier.}
\label{fig:eduardo}
\end{center}
\vspace*{-0.5cm}
\end{figure} 
In Ref.\,\cite{SNR}, S. Narison and E. de Rafael (referred hereafter as SNR) \,\footnote{The main idea comes from Prof. E. de Rafael as I was a student in that time.} have remarked that working with the renormalization group resummed QCD expression
including radiative corrections, the sum rule has instead a Laplace transform property from which all QCD expressions can be derived. 

The Laplace transform sum rule (LSR)  can be derived from the dispersion relation (Hilbert transform) in Eq.\,\ref{eq:2-point} from the useful formulae  (see e.g. the Appendix of \,\cite{SNB1,SNB2})\,:

\bea
{\cal L}\Bigg{[}\frac{1}{(Q^2+m^2)^{\alpha}}\Bigg{]}&=&\frac{\tau^{\alpha}}{\Gamma(\alpha)}e^{-m^2\tau},\nnb\\
{\cal L}\Bigg{[}\frac{1}{(Q^2)^{\alpha}}{\rm ln}\frac{Q^2}{\nu^2}\Bigg{]}&=&\frac{\tau^{\alpha}}{\Gamma(\alpha)}\Big{[}-{\rm ln}\,\tau\nu^2+\psi(\alpha)\Big{]},\nnb\\
{\cal L}\Bigg{[}\frac{1}{(Q^2)^{\alpha}}{\rm ln}^2\frac{Q^2}{\nu^2}\Bigg{]}&=&\frac{\tau^{\alpha}}{\Gamma(\alpha)}\Big{[} {\rm ln}^2\,\tau\nu^2-2\psi(\alpha){\rm ln}\,\tau\nu^2+\psi^2(\alpha)-\psi'(\alpha)\Big{]},\nnb\\
{\cal L}\Bigg{[}\frac{1}{(x)^{\alpha}} \frac{1}{({\rm ln}\,x)^{\beta}}\Bigg{]}&=&y\,\mu(y,\beta-1,\alpha-1),\nnb\\
&\stackrel{y\to 0}{\simeq}&\frac{y^{\alpha+1}}{\Gamma(\alpha)}\frac{1}{(-{\rm ln\,}y)^{\beta}}\Bigg{[}1+\beta\,\psi(\alpha)\frac{1}{{\rm ln\,}y}+{\cal O}\ga \frac{1}{{\rm ln^2}y}\dr\Bigg{]}~~{\rm for}~~~y\equiv \tau\Lambda^2\nnb\\
{\cal L}\Bigg{[}  \frac{{\rm ln\,ln}\, x}  {x^\alpha ({\rm ln}\, x)^\beta} \Bigg{]}& 
\stackrel{y\to 0}{\simeq}
& \frac{y^{\alpha+1}}{\Gamma(\alpha)} \frac{{\rm ln\,ln}\, y}  { ({\rm -ln}\, y)^\beta} \Bigg{[}1+\beta\,\psi(\alpha)\frac{1}{{\rm ln\,}y}+{\cal O}\ga \frac{1}{{\rm ln^2}y}\dr\Bigg{]}
\eea
where ${\cal L}$ is the (inverse) Laplace transform operator  and :
\bea
\psi(z)&\equiv& \frac{d}{dz}{\rm ln}\Gamma(z)~,\nnb\\
\mu(y,\beta,\alpha)&\equiv&\int_0^\infty dx\,\frac{y^{\alpha+x}x^\beta}{\Gamma(\beta+1)\,\Gamma(\alpha+x+1)},\nnb\\
\mu(y,-\beta,\alpha)&=&(-1)^{\beta-1}\frac{d^{\beta-1}}{(dx)^{\beta-1}}\Bigg{[}
\frac{y^{\alpha-x}}{\Gamma(\alpha+x+1)}\Bigg{]}_{x=0}.
\eea
with the useful properties:
\bea
\psi(1)&=&-\gamma_e=-0.5772...~~~~ ({\rm Euler ~constant}),\nnb\\
\mu(y,-1,\alpha)&=&\frac{y^\alpha}{\Gamma(\alpha+1)},\nnb\\
\mu(y,-2,\alpha)&=&\frac{y^\alpha}{\Gamma(\alpha+1)}{[} -{\rm ln}\,y+\psi(\alpha+1){]},\nnb\\
\mu(y,-3,\alpha)&=&\frac{y^\alpha}{\Gamma(\alpha+1)}{[} -{\rm ln}^2\,y-2\psi(\alpha+1){\rm ln}\,y+\psi^2(\alpha+1)-\psi'(\alpha+1){]}.
\eea
In addition to previous formulae, the expression of the integral:
\beq
\int_0^{t_c}dt\,t^n\,e^{-t\tau}=(n-1)!\,\tau^{-n}(1-\rho^n)~~~~{\rm with~~~~}
\rho^n=e^{-t_c\tau}\ga 1+t_c\tau+\cdots \frac{(t_c\tau)^n}{n!}\dr
\eeq
is useful for the treatment of the QCD continuum contribution to the LSR from the QCD spectral  function as parametrized in Eq.\,\ref{eq:duality}.

\section{The QCD expressions of the prototype $\rho$ meson and pion LSR}
We illustrate the derivation of the Laplace sum rule in the case of the vector and pseudoscalar (divergence of the axial-vector) currents:
\bea
V_\mu(x)&=&\bar u(\gamma_\mu)d (x),\nnb\\
\partial_\mu A^\mu(x)&=&(m_u+m_d)\bar u(i\gamma_5)d (x)
\eea
and the corresponding two-point functions $\Pi(q^2)$ and $\psi_5(q^2)$ 
used in Ref.\,\cite{SNR} for demonstrating (for the first time) the Laplace transform properties of the SVZ sum rule\,\footnote{In these pedagogical examples, we shall limit to the perturbative QCD expression to order $\alpha_s^2$. Analysis including higher order terms can be found in\,\cite{SNe23,SNLIGHT}.} :
\bea
-(g_{\mu\nu}q^2-q_\mu q_\nu)\Pi(q^2)&=&i\int d^4x\, e^{iqx}\la 0\vert {\cal T}\, V_\mu(x) V^\dagger_\nu(0)\vert  0\ra,\nnb\\
\psi_5(q^2)&=&i\int d^4x\, e^{iqx}\la 0\vert {\cal T}\, \partial^\mu A_\mu(x) \ga\partial^\mu A_\mu\dr^\dagger(0)\vert  0\ra
\eea
\subsection{The $\rho$ meson channel}
As the corresponding two-point correlator is once substracted, it is convenient to work with its first derivative:
\beq
\chi^{(1)}(Q^2)\equiv \ga-\frac{d}{dQ^2}\dr \Pi(Q^2) = \frac{C_0}{Q^2}+\frac{C_2\la O_2\ra}{(Q^2)^2}+2\frac{C_4\la O_4\ra}{(Q^2)^3}+3\frac{C_6\la O_6\ra}{(Q^2)^4}+\cdots
\eeq
with (see e.g.\,\cite{SNB1,SNB2}):
\bea
C_0&=&\frac{1}{4\pi^2}\big{[}1+ a_s + \bar a_s^2R_2+\cdots\big{]},\nnb\\
C_2\la O_2\ra&=&3(m_u^2+m_d^2),\nnb\\
C_4\la O_4\ra&=&4\pi^2\big{[}m_u\la \bar \psi_u\psi_u\ra+m_d\la \bar \psi_d\psi_d\ra\big{]}+\frac{\pi}{3}\la\alpha_s G^2\ra +{\cal O}(m^4_i\,{\rm ln} (m_i^2/q^2),\nnb\\
C_6\la O_6\ra&=&\frac{896}{81}\pi^3 \rho\,\alpha_s\la\bar \psi_q\psi_q\ra^2,
\eea
where $R_2=1.986-0.115\,n_f$ for $n_f$ flavours.  $\chi^{(1)}(Q^2)$
 is superconvergent and obeys the homogeneous Renormalization Group Equation (RGE):
\beq
\ga-Q^2\frac{\partial}{\partial Q^2}+\beta(\alpha_s)\alpha_s\frac{\partial}{\partial\alpha_s}-\sum_i[1+\gamma(\alpha_s)] x_i\frac{\partial}{\partial x_i}\dr \chi^{(1)}(Q^2),
\label{eq:rge}
\eeq
with the running parameters to order $\alpha_s$ are (see e.g.\cite{FNR,RUNDEC,SNB1,SNB2}):
\bea
 a_s&\equiv& \frac{\bar \alpha_s(Q^2)}{\pi} = a_s^{(0)}\Big{[}1-a_s^{(0)}\frac{\beta_2}{\beta_1}\,{\rm ln\, ln}(Q^2/\Lambda^2)\Big{]} \nnb\\
\bar m_q(Q^2)&\equiv& x_i\sqrt{Q^2}\equiv \frac{\hat m_q}{(-\beta_1 a_s)^{\gamma_1/\beta_1)}}\Big{[} 1+\frac{\beta_2}{\beta_1}\ga \frac{\gamma_1}{\beta_1}-\frac{\gamma_2}{\beta_2}\dr a_s+2.707a_s^2\Big{]},
\label{eq:run}
\eea
where:
\beq
a_s^{(0)}=\frac{2}{-\beta_1\,{\rm ln}(Q^2/\Lambda^2)}~~ :  ~~~~~~~~~~~~\Lambda= (342\pm8)~{\rm MeV} ~{\rm for~3~flavours}.
\eeq
$\beta_{1,2}$ and $\gamma_{1,2}$ are respectively the 1st and 2nd coefficients of the $\beta$ function and mass anomalous dimension. For $n_f$ quark flavours, they read:
\beq
\beta_1=-\frac{11}{2}+\frac{1}{3}n_f,~~~~~~\beta_2=-\frac{51}{4}+\frac{19}{12}n_f, ~~~~~~\gamma_1=2,~~~~~ ~\gamma_2=\frac{101}{12}-\frac{5}{18}n_f. 
\eeq
Using the previous formulae of the Laplace transform into the QCD expressions of the corresponding two-point correlator, one can deduce the Laplace transform\,\cite{SNR,SNB1,SNB2}:
\bea
{\cal L}^V_0&\equiv& \int_0^\infty e^{-t\tau}\frac{1}{\pi}{\rm Im}\Pi_V(t)= \tau^{_1}\ga\frac{1}{4\pi^2}\dr\Bigg{\{}1+a_s(\tau)+a_s^2(\tau)\big{[}R_2-\frac{1}{2}\beta_1\gamma_E+(\beta_2/\beta_1){\rm ln\,ln}(\tau\Lambda^2)\big{]}\nnb\\
&&+4\pi^2\tau^2\big{[}m_u\la \bar \psi_u\psi_u\ra+m_d\la \bar \psi_d\psi_d\ra\big{]}+\frac{\pi}{3}\tau^2\la\alpha_s G^2\ra + \tau^3 \ga\frac{1}{2}\dr\frac{896}{81}\pi^3 \rho\,\alpha_s\la\bar \psi_q\psi_q\ra^2\Bigg{\}},
\eea
where the factor $\rho=1$ if one uses the vacuum saturation assumption for the four-quark condensate. 
\subsection{The pion channel}
The corresponding two-point function behaves as $(q^2)^2\, {\rm ln}(-q^2/\nu^2)$and is then twice substracted. Thererefore, its 2nd derivative $\psi_5^{(2)}(q^2)$ is superficially convergent and obeys
the RGE in Eq.\,\ref{eq:rge}. Taking the Laplace transform of $\psi_5^{(2)}$, one obtains:
\cite{SNR,SNB1,SNB2}\,\footnote{For an updated expression including higher order contributions, see e.g.\,\cite{SNLIGHT}.}:
\bea
{\cal L}_0^{\pi}&=&\frac{3}{8\pi^2}\frac{(\hat m_u+\hat m_d)^2}
{\ga-\rm{ln}\,\sqrt{\tau}\Lambda\dr^{2\gamma_1/-\beta_1}}\tau^{-2}
\Bigg{\{}1-\big{[} \bar m_u^2+\bar m_d^2+(\bar m_u-\bar m_d)^2\big{]}\tau\nnb\\
&&+\,a_s\Bigg{[} \frac{11}{3}+2\gamma_E+ \frac{2}{-\beta_1}\ga \gamma_2-\gamma_1\frac{\beta_2}{\beta_1}\dr +2\gamma_1\frac{\beta_2 }{\beta_1^2}\,{\rm ln\,\ga -ln\,\tau\Lambda^2\dr}       \big{]}  \Bigg{]}\nnb\\
&&-\,\frac{8}{3}\pi^2\tau^2\Bigg{[} \ga m_d-\frac{m_u}{2}\dr \la\bar \psi_u\psi_u\ra 
+ \ga m_u-\frac{m_d}{2}\dr \la\bar \psi_d\psi_d\ra \Bigg{]}+\frac{\pi}{3}\tau^2\la\alpha_s G^2\ra\nnb\\
&&-\frac{4\pi^2}{3}\tau^3\Big{[} m_d\la\bar \psi_uG\psi_u\ra  +
\frac{32}{27}\pi \rho\alpha_s\ga \la\bar \psi_u\psi_u\ra^2+\la\bar \psi_d\psi_d\ra^2-9 
\la\bar \psi_u\psi_u\ra \la\bar \psi_d\psi_d\ra    \dr\Big{]}~
\Bigg{\}}.
\eea
\section {Optimization procedure for the LSR}
\subsection{The SVZ sum rule window}
The second important step in the sum rule analysis is the extraction of the optimal result from the sum rule as, in principle, the Laplace sum rule (LSR) variable $\tau$, the degree of moments $n$ and the QCD continuum threshold $t_c$ are free external variables. In their original work,  SVZ have adjusted the values of $M^2\equiv 1/\tau$ and $t_c$ using some guessed $\%$ contributions, for finding the sum rule window where the QCD continuum contribution is less than some input number while the ground state one is bigger than some input number and where,  in this sum rule window, the OPE is expected to converge.  The arbitrariness values of these numbers have created some doubts for non-experts on the results from the sum rules, in addition to the ones on the eventual non-correctness of the non-trivial QCD-OPE expressions. Unfortunately, many sum rules practitioners continue to use this inaccurate SVZ criterion. 
\subsection{$\tau$-stability from quantum mechanics and $J/\Psi$}
\begin{figure}[H]
\begin{center}
\includegraphics[width=11.cm]{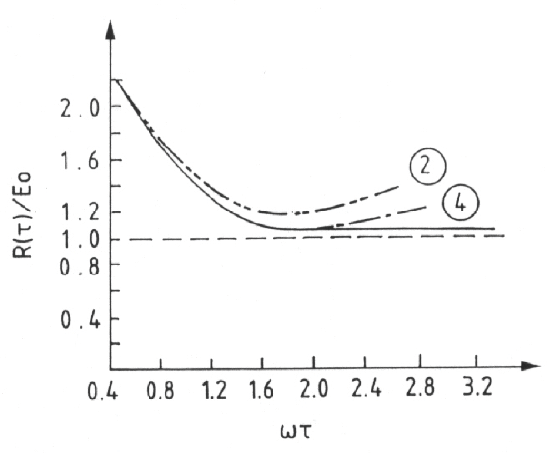}
\caption{\normalsize Behaviour of the ground state mass versus the time variable $\tau$ for different truncation of the series from Ref.\,\cite{BELLa,BERTa}. The horizontal line is the exact solution. $\omega$ is the harmonic oscillator frequency.}
\label{fig:oscillo1}
\end{center}
\end{figure} 

\begin{figure}[hbt]
\begin{center}
\includegraphics[width=11.cm]{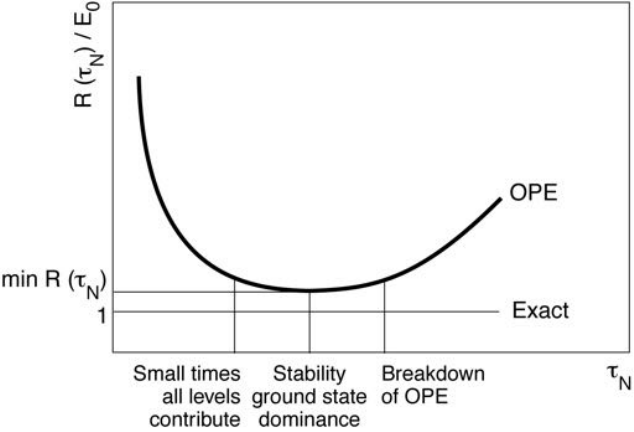}
\caption{\normalsize  Schematic behaviour of the $J/\psi$ mass  versus the sum rule variable $\tau_N$ from Ref.\,\cite{BELLa,BERTa}. The horizontal line is the experimental mass.} \label{fig:oscillo2}
\end{center}
\end{figure} 
\subsection{Examples of $\tau$-stability from the $\Upsilon$ and $B$ mesons}

\begin{figure}[H]
\begin{center}
\includegraphics[width=13.5cm]{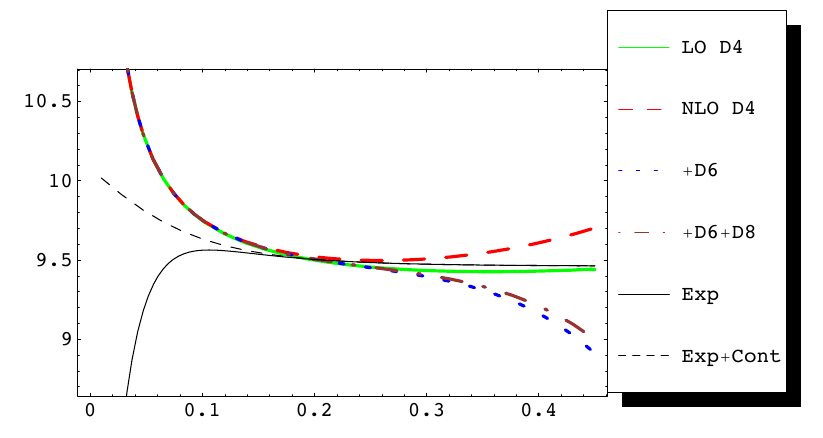}
\caption{\normalsize  Behaviour of the ratio of moments $\sqrt{{\cal R}^b_{10}}(\overline{m}_b^2)$ in GeV versus $\tau$ in GeV$^{-2}$ and for $\overline{m}_b(\overline{m}_b) = 4212$ MeV from\,\cite{SNH12}. The black continuous (resp. short dashed) curves are the experimental contribution including (resp. without) the QCD continuum (it is about the $M_{\Upsilon}$). The green (thick continuous) is the PT contribution including the $D=4$ condensate to LO. The long dashed (red) curve is the contribution including the $\alpha_s$ correction to the $D=4$ contribution. The short dashed (blue) curve is the QCD expression including the $D=6$ condensate and the dot-dashed (red-wine) is the QCD expression including the $D=8$ contribution. } 
\label{fig:ratiob}
\end{center}
\vspace*{-0.5cm}
\end{figure} 
\nin
\begin{center}
{\begin{figure}[hbt]
\begin{center}
{\includegraphics[width=13.5cm]{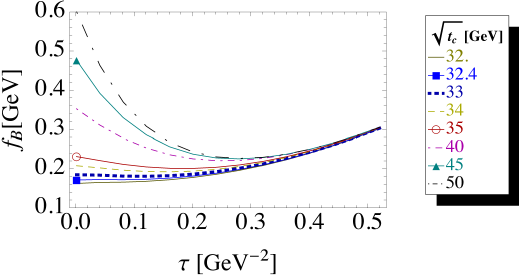}}
\end{center}
\caption{\normalsize Behaviour of $f_B$ from Ref.\,\cite{SNFB12} versus the sum rule variable $\tau$ for different values of the QCD continuum threshold $t_c$ and for a given $\mu=$ 3 GeV and $\overline{m}_b(\overline{m}_b) = 4177$ MeV.}
\label{fig:fb}
\end{figure}}
\end{center}
Hopefully, in the examples of harmonic oscillator in quantum mechanics and of the non-relativistic charmonium sum rules, Refs.\,\cite{BELLa,BERTa} have shown that the optimal information 
from the analysis for a truncated series is obtained at the minimum or inflexion point in $\tau$ of the approximate theoretical curves (see Figs\,\ref{fig:oscillo1} and \ref{fig:oscillo2}). This optimal criterion corresponds  to the {\it (principle of minimum sensitivity of the physical parameters on the external sum rule variable $\tau$)} and where the exact solution is reached when one adds more and more terms in the approximate series.  

In the case of hadrons, we illustrate in Figs.\,\ref{fig:ratiob} and \ref{fig:fb} the  analysis for the $\Upsilon$ systems and for the $B$ meson decay constant using relativistic  sum rules. 

\subsection{$t_c$ and $\mu$-stabilities}
Later on, we have extended this $\tau$-stability criterion to the continuum threshold variable $t_c$\,\footnote{In many papers, the optimal value is extracted at the lowest value of $t_c$ ! but the result still increases with the $t_c$ changes.  } and to the arbitrary Perturbative (PT) subtraction point $\mu$\,\footnote{One can also use the RGE resummed solution which is equivalent to take $\mu^2=1/\tau$ but in some case the value of $\tau$ is relatively large such that the OPE is not well behaved. A such choice of $\mu$ value is often outside the $\mu$-stability region.}. 
The lowest value of $ t_c$ corresponds to the beginning  of $\tau$-stability while the $t_c$-stability 
corresponds to a complete lowest ground state dominance in the spectral integral. In our analysis, we always consider the conservative optimal result inside this $t_c$-region\,\footnote{$\sqrt{t_c}$ is often identified to the mass of the 1st radial excitation which is a crude approximation as the QCD continuum smears all higher state contributions to the spectral function.}. 
\subsection{Ground state versus the QCD continuum}
For some low value of $(t_c$ and large value of $\tau$, one can have some flat curves or some (apparent) minimum in $\tau$. Then, to check or/and to restrict the optimal region in this case, one also  requests that the contribution of the ground state to e.g. the spectral integral (e.g. moment sum rule) is larger than the QCD continuum one. One can formulate this constraint in a more rigorous way as (see e.g.\,\cite{SNPC} for some examples of applications)\,:
\beq
\hspace*{-0.75cm}R_{P/C} \equiv\frac{\int_{t>}^{t_c}dt\,e^{-t\tau}{\,\rm Im}\,\Pi(t)}{\int_{t_c}^{\infty}dt\,e^{-t\tau}{\,\rm Im}\,\Pi(t)}\geq 1.
\eeq
\vspace*{-0.35cm}
\section{Phenomenology of the $\rho$-meson LSR}
\subsection{Lower bound on the $\rho$-meson coupling from the lowest moment ${\cal L}_0$}
We use the MDA parametrization of the spectral function and extract the $\rho$-meson coupling\,\footnote{Analysis using the complete $e^+e^-\to I=1$ Hadrons data can e.g. be found in Refs.\,\cite{LNT,SNe23}}.  :
\beq
\frac{1}{\pi} \mbox{Im}\Pi_\rho(t) =\frac{M_\rho^2}{2\gamma_\rho^2}\,~~~~~~~~~~~{\rm with}\,~~~~~~~~~~~
\Gamma_{\rho\to e^+e^-}=\frac{2}{3}\alpha^2\pi\frac{M_\rho}{2\gamma_\rho^2}\,  ~~~~~~~~~~~ {\rm and} ~~~~~~~~~~~\gamma_\rho^{exp}= 2.479(15)
\eeq

\begin{figure}[hbt]
\begin{center}
\includegraphics[width=17.5cm]{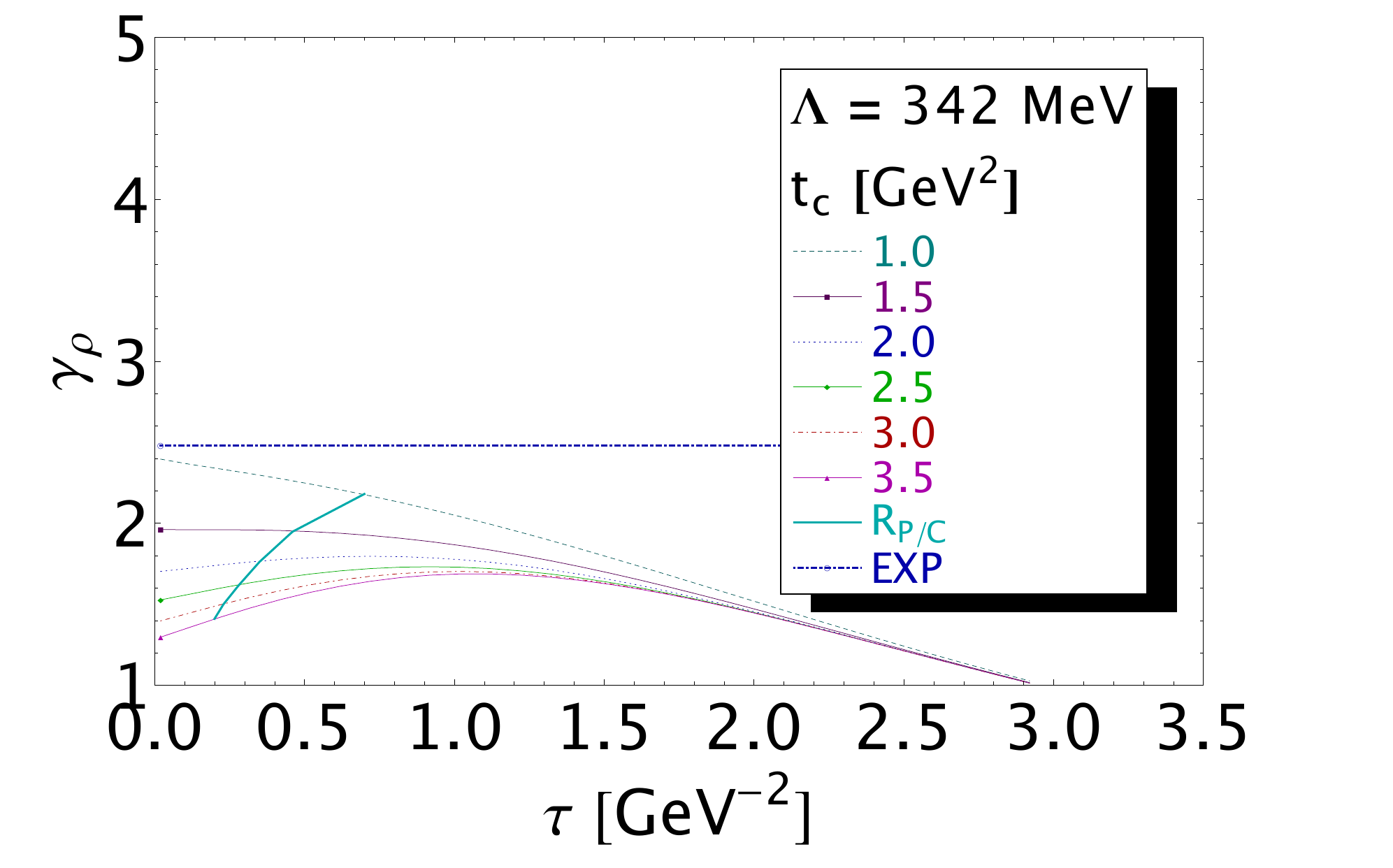}
\caption{\normalsize  Lower bound  of the $\gamma_\rho$ coupling  versus the sum rule variable $\tau$ for different values of the continuum threshold $t_c$.} \label{fig:gamrho}
\end{center}
\vspace*{-0.5cm}
\end{figure}

\d Using positivity of the spectral integral and evaluating the LSR at $\tau=1/M_{\rho}^2$,  SNR deduces the inequality to leading order of PT series\,\footnote{Notice that this inequality has been considered as an estimate in the original SVZ paper\,\cite{SVZa}.}:
\beq
\gamma_\rho^2\geq \frac{4\pi}{2\,e}\Big{[}1+{\cal O}(\alpha_s)-\frac{\pi}{3}\frac{\la \alpha_s G^2\ra}{M_\rho^4}\Big{]}\lrar \gamma_\rho\geq  2.43.
\eeq
We improve this bound by adding higher order terms in the PT series and including the contribution of the condensates up to dimension-six.  We show the analysis in Fig.\ref{fig:gamrho} for different values of $t_c$ and $\tau$.  The $R_{P/C}$ condition is shown by the cyan curve where the lowest ground state dominates in the delimited right region.  Optimal results are obtained for the sets $(\tau,t_c)$  from (0.5,1.5) to (1.1,3.0) in units of (GeV$^{-2}$, GeV$^2$). This leads to the improved lower bound at order $\alpha_s^2$\,:
\beq
\gamma_\rho \geq 1.83(13)_{t_c}(1)_{a_s}(1)_{G^2}(1)_{\bar\psi\psi^2}.
\eeq
\vspace*{-1cm}
\vspace*{-0.5cm}
\begin{figure}[hbt]
\begin{center}
\includegraphics[width=17.5cm]{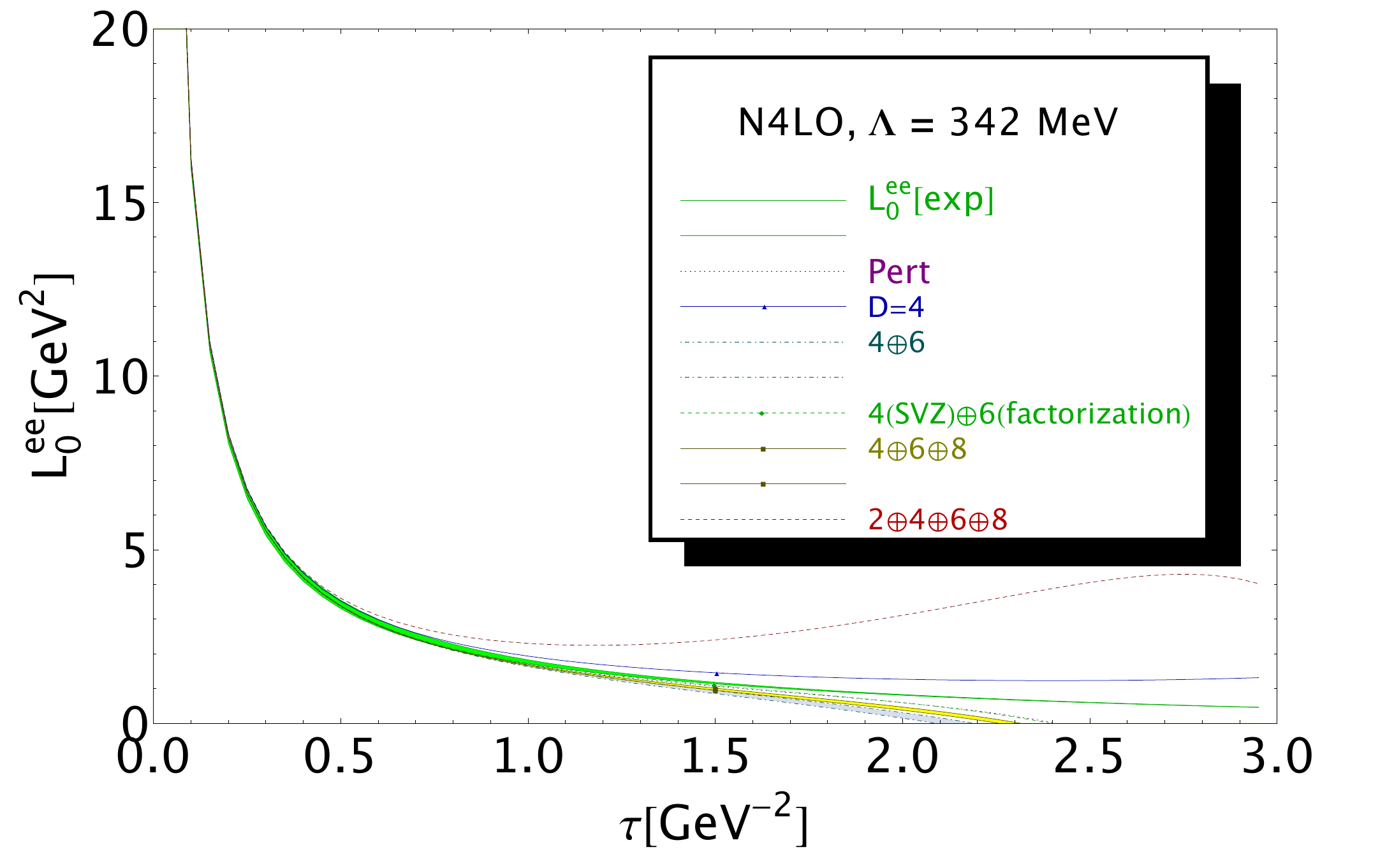}
\caption{\normalsize  $\tau$-behaviour of the lowest moment ${\cal L}_0$ for different truncation of the OPE.}\label{fig:Lee}
\end{center}
\vspace*{-0.5cm}
\end{figure} 
\subsection{The lowest moment ${\cal L}_0$ using the $e^+e^-\to I=1$ Hadrons data}

It is informative to compare the QCD prediction of ${\cal L}_0$ with the one where the
complete $e^+e^-\to I=1$ Hadrons data is used to estimate the spectral integral. This analysis has been done recently  in Ref.\,\cite{SNe23} which we show in Fig.\,\ref{fig:Lee}

One can observe that there is a good agreement with the experimental $e^+e^-$ data. 
and the theory prediction using the previous sets of QCD parameters up to $D=6,8$ condensates.

\subsection{$M_\rho$ from the lowest ratio of moments ${\cal R}_{10}$}
\begin{figure}[hbt]
\begin{center}
\includegraphics[width=17.5cm]{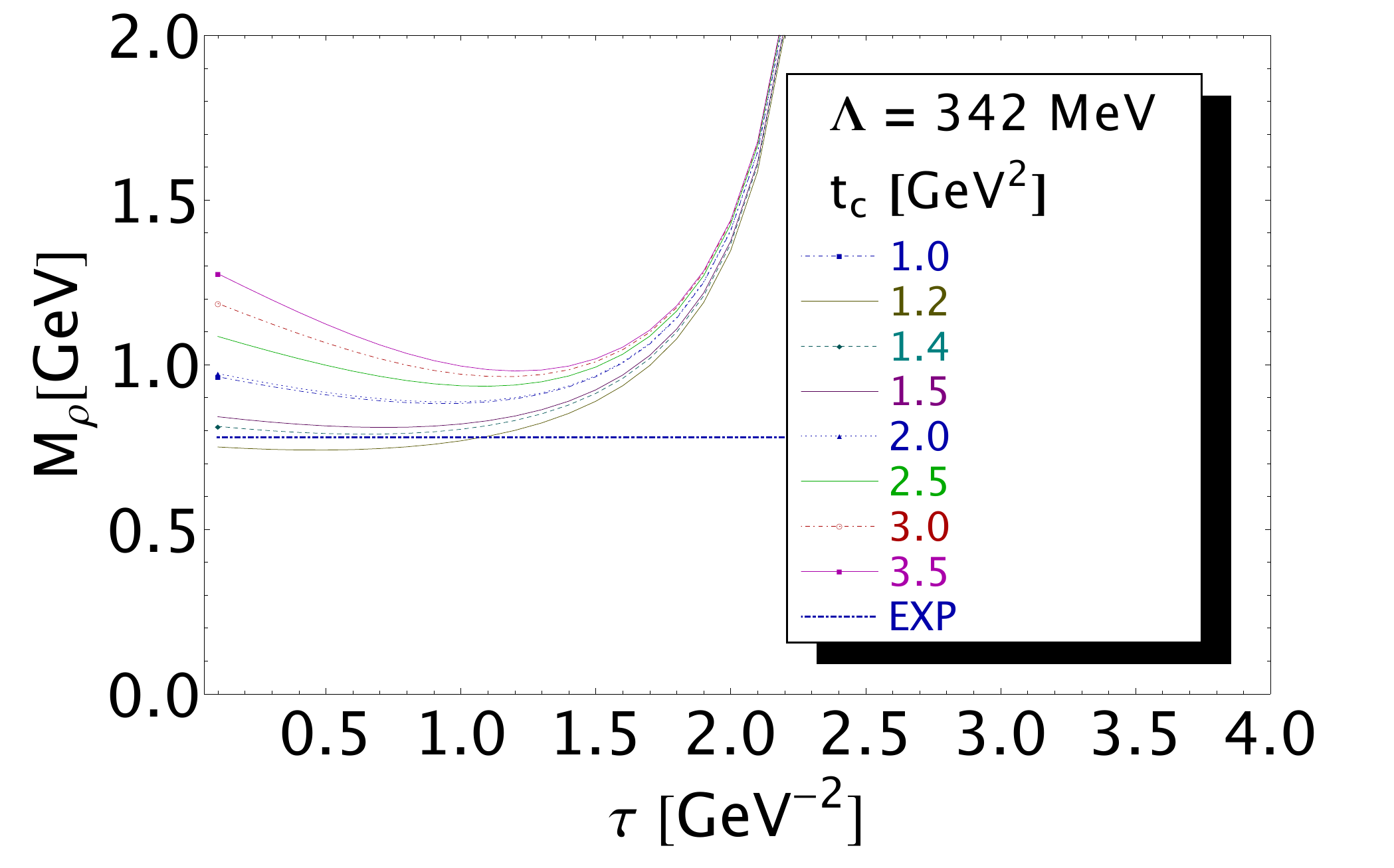}
\caption{\normalsize   $M_\rho$  versus the sum rule variable $\tau$ for different values of the continuum threshold $t_c$.} \label{fig:mrho}
\end{center}
\vspace*{-0.5cm}
\end{figure}

\d Within MDA, one expects that the ratio of moments provides the mass squared of the 
$\rho$-meson ground state. We show the analysis versus $\tau$ and for different values of $t_c$ in Fig.\,\ref{fig:mrho}.
The optimal results are obtained for the sets $(\tau,t_c)$ from (0.5,1.2) to (1.2,3.0)
in units of (GeV$^{-2}$, GeV$^2$). We obtain:
\beq
M_\rho\simeq 848(109)_{t_c}(3)_{a_s}(3)_{G^2}(13)_{\bar\psi\psi^2},
\eeq
where the error comes mainly from the value of $t_c$ which we have taken in the conservative range from the beginning of $\tau$ stability until the $t_c$-one. 

\d One can restrict this range of $t_c$ value using the constraint from the first moment of FESR. It reads\,\cite{FESR1,FESR2}:
\beq
\frac{M_\rho^2}{2\gamma_\rho^2} = \frac{t_c}{4\pi^2}\Big{[} 1+a_s(t_c)+a_s^2(t_c)\big{[}R_2-\frac{\beta_1}{2}-(\beta_2/\beta_1){\rm ln\,ln}(tc/\Lambda^2)\big{]}\Big{]},
\eeq
from which we deduce:
\beq
t_c\simeq 1.962(24)_{t_c}(2)_{\alpha_s}~{\rm GeV}^2. 
\label{eq:tc}
\eeq

\d One should note that the low value of $t_c$ often chosen by many ``QCD sum rules practitioners" to reproduce the value of the experimental mass or to extract the mass of the ground state does not coi ncide with the duality constraint from FESR.  At the value of $t_c$ in Eq.\,\ref{eq:tc}, we obtain from the ratio of moments\,:
\beq
M_{\rho}\simeq 881~{\rm MeV}.
\eeq

\d From the previous analysis, the ratio of moments reproduces within 10\% the experimental meson mass. 
\subsection{QCD condensates from ${\cal R}_{10}$ using $e^+e^-\to$ Hadrons data}
\d The ratio of moments has the advantage to be less sensitive to $\alpha_s$ correction than the moment  ${\cal L}_{0}$ as the PT correction starts to order $\alpha_s^2$. It is then expected to be a good place for extracting the QCD condensates. This analysis has been initiated by Ref.\,\cite{LNT} and revised recently in Ref.\,\cite{SNe23}. Using as input the value of the gluon condensate obtained from heavy quark sum rules\,\cite{SNparam,SNREV22,SNREV21} given in Eq.\,\ref{eq:g2}, one obtains:
\beq
\rho \alpha_s\la\bar\psi\psi\ra^2\simeq 5.98(64)10^{-4}\,{\rm GeV}^6,~~~~~~ ~~~~~~~~~C_8\la O_8\ra=(4.3\pm 3.0)\times 10^{-2}\,{\rm GeV}^8`, 
\eeq
where the value of the four-quark condensate is in good agreement with the one in Eq.\,\ref{eq:d6}. 

\begin{figure}[hbt]
\begin{center}
\includegraphics[width=17.5cm]{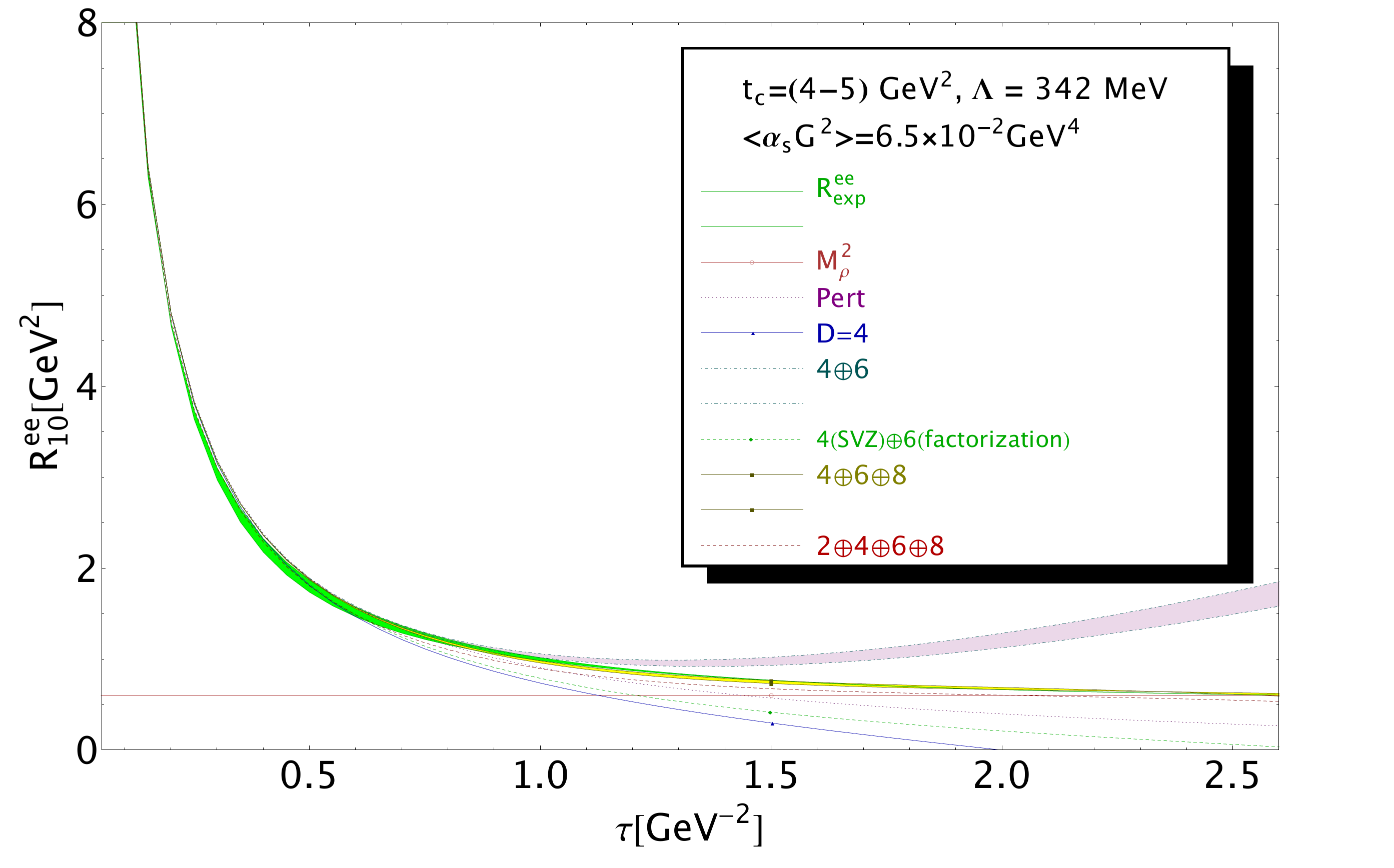}
\caption{\normalsize  $\tau$-behaviour of the ratio of moments ${\cal R}_{10}$ for different truncation of the OPE.}\label{fig:Ree}
\end{center}
\vspace*{-0.5cm}
\end{figure}

\d We show in Fig.\,\ref{fig:Ree} a comparison of the experimental expression of ${\cal R}_{10}$ with the QCD prediction for different truncations of the OPE. One can notice a good agreement when the condensates up to dimension 6,8 are included. 
\section{Sum of light quark masses from the pion LSR}
\subsection{Lower bound on $(\hat{m}_u +\hat{m}_d)$}
\begin{figure}[hbt]
\begin{center}
\includegraphics[width=17.5cm]{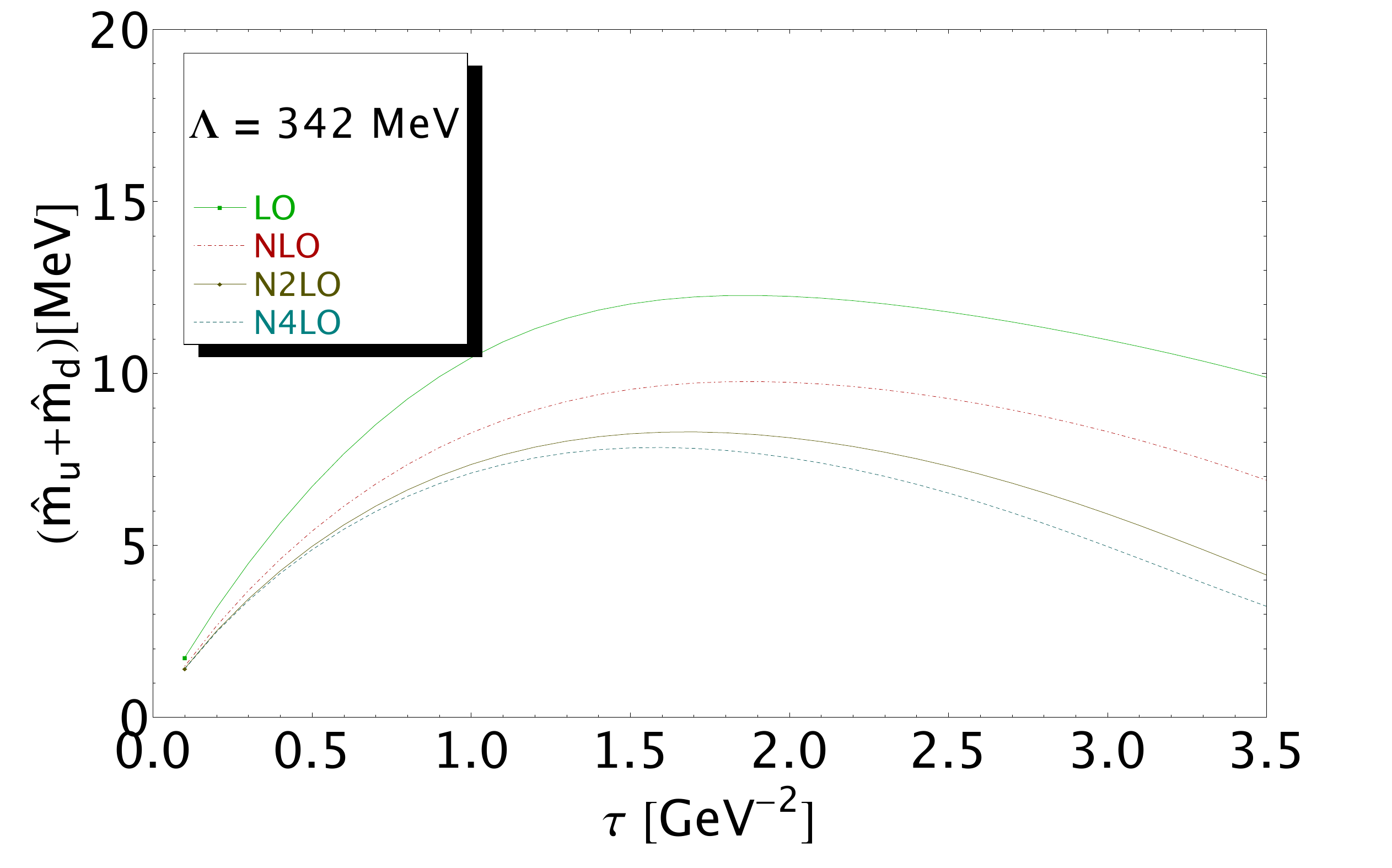}
\caption{\normalsize  $\tau$-behaviour of RGI $(\hat{m}_u +\hat{m}_d)$ lower bound for different truncation of the PT series.}\label{fig:mud}
\end{center}
\vspace*{-0.5cm}
\end{figure}

Using the pion sum rule, SNR has derived a lower bound on the RGI quark mass defined in Eq.\,\ref{eq:run} from the positivity of the spectral function and retaining the pion contribution.  The results have been given for different values of the QCD scale $\Lambda$. Given the present progress on the determination of $\Lambda$, we show the new version of the figure given in Ref.\,\cite{SNR} in  Fig.\,\ref{fig:mud} for different truncation of the PT series. 
We deduce the optimal lower bound at the maximum of the curves\,:
\bea
(\hat{m}_u +\hat{m}_d) &\geq& 10.13(6)_{\alpha_s}(22)_{G^2} (37)_{\bar\psi\psi^2}~{\rm MeV} = 10.13(44)~{\rm MeV}~~~:~~~~{\rm NLO},\nnb\\
&\geq& 7.92(26)_{\alpha_s}(1)_{G^2} (8)_{\bar\psi\psi^2} ~{\rm MeV}~~~= 7.92(27)~{\rm MeV}~~~~:~~~~{\rm N4LO},
\eea
This bound can be improved by introducing the contribution of the next radial excitation $\pi(1300)$\,\cite{SNLIGHT} and/or some alternative approaches\,\cite{LELLOUCH}.

\subsection{Estimate of $(\hat{m}_u +\hat{m}_d)$}
One can transform the previous lower bound by introducing the $\pi(1300)\oplus$ QCD continuum for parametrizing the spectral function. Using the standard OPE, one obtains to order $\alpha_s^4$ for the RGI and running masses $\overline{m}_q$ evaluated at 2 GeV\,\cite{SNLIGHT}\,\footnote{Extension of the analysis to some other channels are discussed in this paper.}:
\beq
(\hat{m}_u +\hat{m}_d) =9.76(88)~{\rm MeV}~~~\to~~~(\overline{m}_u +\overline{m}_d) (2)=8.08(61)~{\rm MeV},
\eeq
where one can notice that the main uncertainties and the size of the central value come from the parametrization of the spectral function (radial excitations $\oplus$ choice of the QCD continuum threshold $t_c$). The error due to the way for truncating  the PT series only affects slightly  the total error (see e.g.\,\cite{KAHN}). 
These results agree with lattice calculations\,\cite{LATTLIGHT} and from some other approaches quoted in PDG\,\cite{PDG}.
\section{Some other applications of LSR}
These applications have been already summarized in the recent short reviews\,\cite{SNREV21,SNREV22} where  orginal references and some dedicated reviews can be found. They concern the\,:

\b Light baryons $qqq'$ ligth quark states 

\b Heavy quarks $\bar QQ$ sector.

\b Heavy-light $\bar Qq$ states.

\b So-called exotic states :

\hspace*{0.5cm} \d  Glueballs / gluonia $gg, ggg$ states.

\hspace*{0.5cm} \d  Hybrids $\bar q g q, \cdots$ states.

\hspace*{0.5cm} \d  Light $(\bar q\bar q') (qq')$, heavy-light  $(\bar q\bar Q) (Qq)\cdots$ , fully-heavy $(\bar Q\bar Q) (QQ),\cdots$ four-quark states.

\hspace*{0.5cm} \d  Light $(\bar q q') (q\bar q')$, heavy-light  $(\bar q Q) (Q\bar q)\cdots$ , fully-heavy $(\bar Q Q) (\bar QQ),\cdots$ molecule states.


We plan to develop these different parts in a more extended review on QCD spectral sum rules. 

\section*{Concluding remarks}
LSR and (in general) QCD spectral sum rules (QSSR) are useful tools to tackle within a good approximation the properties of hadrons and for extracting the fundamental  parameters of the QCD Lagrangian. If properly done, QSSR is a serious alternative or/and a supplement to the lattice QCD numerical simulations. 


\end{document}